\documentclass[12pt]{article}
\usepackage{xcolor}
\usepackage{multirow}
\usepackage{comment}
\usepackage{amsmath}
\usepackage{amssymb}
\usepackage{graphicx}
\usepackage{here}
\usepackage{subcaption}
\usepackage{url}
\usepackage[sort&compress, numbers, merge]{natbib}
\usepackage{tcolorbox}
\usepackage{bbm}
\usepackage{physics}% Added by SK
\usepackage[compat=1.1.0]{tikz-feynhand}
\usepackage{slashed}
\usepackage{mathtools}
\usepackage{comment}
\usetikzlibrary{calc,arrows.meta,decorations.markings}

\usepackage[subpreambles=true]{standalone} 
\usepackage{todonotes}

\numberwithin{equation}{section}

\newcommand{\MSB}{\overline{\mathrm{MS}}}
\def\Umt{\mathrm{U}(1)_{L_{\mu}-L_{\tau}}}
\def\calL{\mathcal{L}}

\def\Nbar{\bar{N}}

\def\ebar{\bar{E}}
\def\s12{s_{12}}

\usepackage[colorlinks=true, linkcolor=blue, citecolor=blue,
urlcolor=black]{hyperref}

\begin{document}

\begin{titlepage}

\begin{flushright}
IPMU26-0028
\end{flushright}

\vskip 1.1cm

\begin{center}

{\Large \bf 
Radiative Breaking of Two-Zero Neutrino Mass Minors:
Revisiting the $\boldsymbol{\Umt}$ Model
}

\vskip 1.2cm
Masahiro Ibe$^{a,b}$,
Jun Miyamoto$^{a}$ and
Satoshi Shirai$^{b}$ 
\vskip 0.5cm

{\it

$^a$ {ICRR, The University of Tokyo, Kashiwa, Chiba 277-8582, Japan}

$^b$ {Kavli Institute for the Physics and Mathematics of the Universe
(WPI), \\The University of Tokyo Institutes for Advanced Study, \\ The
University of Tokyo, Kashiwa 277-8583, Japan}

}

\vskip 1.0cm

\begin{abstract}
The two-zero minor structure predicted by flavor symmetries is usually discussed as a tree-level relation among low-energy neutrino parameters. 
We point out that this relation can be significantly modified by radiative corrections even when the two-zero minor structure is enforced by an underlying symmetry at tree level. 
As a concrete example, we analyze the minimal $\Umt$ model and compute the universal one-loop threshold corrections associated with the type-I seesaw sector. 
These corrections generate flavor-dependent contributions to the Weinberg operator and violate the exact two-zero minor conditions once the $\Umt$ symmetry is spontaneously broken. 
This effect relaxes the tree-level lower bound on the total neutrino mass and thereby weakens the tension between the model and cosmological neutrino-mass constraints.
\end{abstract}

\end{center}
\end{titlepage}

\clearpage
\section{Introduction}
The origin of neutrino masses and lepton flavor mixing remains one of the central questions in particle physics.
Among the many possible explanations, the seesaw mechanism provides a particularly simple and compelling framework, in which the smallness of neutrino masses is naturally understood by introducing heavy right-handed neutrinos~\cite{Minkowski:1977sc,Yanagida:1979as,*Yanagida:1979gs,Gell-Mann:1979vob,Glashow:1979nm,Mohapatra:1979ia}.
The seesaw mechanism is, however, not predictive by itself: without a convincing theory of the flavor structure of the Yukawa coupling matrices, the number of free parameters exceeds that of low-energy neutrino observables.
For this reason, many attempts have been made to supplement the seesaw framework with additional assumptions to obtain predictive relations among neutrino masses and mixing parameters.
Widely studied examples include flavor symmetries~(see e.g., Refs.\,\cite{Altarelli:2010gt,Ishimori:2010au,King:2013eh}), 
and texture-zero or minor-zero structures in the neutrino mass matrix~(see e.g., Refs.\,\cite{Frampton:2002yf,Xing:2002ta,Lashin:2007dm,Dev:2009he,Fritzsch:2011qv,Harigaya:2012bw,Singh:2016qcf}).

It should be emphasized, however, that when such parameter relations are realized by flavor symmetries, they are typically tree-level relations.
It is therefore important to examine how robust the resulting predictions are against radiative corrections.
Indeed, Refs.\,\cite{Grimus:2002nk,AristizabalSierra:2011mn} pointed out that, in fine-tuned seesaw models, one-loop corrections to the neutrino mass matrix can be so large that the tree-level predictions may differ significantly from the corresponding one-loop predictions.
For models with texture-zero or minor-zero structures, it is therefore essential to specify the underlying mechanism that realizes these structures; only then can one meaningfully assess whether the resulting predictions are stable under radiative corrections.

In this paper, we discuss the minimal $\Umt$ model~\cite{Foot:1990mn,He:1991qd,Foot:1994vd,Gninenko:2001hx,Baek:2001kca,Murakami:2001cs,Ma:2001md}, which is based on one of the simplest anomaly-free flavor symmetries.
In this model, both the charged-lepton Yukawa coupling and the neutrino Dirac Yukawa coupling are diagonal as a consequence of the symmetry.
In the minimal setup, the $\Umt$ symmetry is broken by a single complex scalar field, leading to a right-handed neutrino Majorana mass matrix whose $(\mu,\mu)$ and $(\tau,\tau)$ entries vanish.
After integrating out the right-handed neutrinos, this structure gives rise to the two-zero minor conditions in the light-neutrino mass matrix at tree level. 
This condition imposes nontrivial 
relations among neutrino masses, 
mixing angles, and CP phases, making the minimal $\Umt$ model highly predictive~\cite{Asai:2017ryy,Asai:2018ocx,Asai:2020qax}.

The predictive power of the two-zero minor structure has recently become especially important in view of the improved neutrino and cosmological data.
Since the two-zero minor condition tends to require a quasi-degenerate neutrino mass spectrum, the minimal $\Umt$ model is strongly constrained by cosmological upper bounds on the sum of neutrino masses, together with neutrino oscillation data, direct neutrino mass measurements, and neutrinoless double-beta decay.
Indeed, a recent global analysis showed that the minimal $\Umt$ model is already significantly disfavored under the assumption of the $\Lambda$CDM cosmology~\cite{Ibe:2025rwk}.

This conclusion, however, relies on the tree-level two-zero minor structure.
In the minimal $\Umt$ model, radiative corrections can violate the two-zero minor structure as the $\Umt$ symmetry is spontaneously broken.
Thus, even if the exact two-zero minor condition is imposed at tree level, the low-energy neutrino mass matrix may deviate from it.
In this paper, we quantify the resulting deviations from the tree-level prediction and examine whether they can relax the tension between the minimal $\Umt$ model and current neutrino and cosmological data.

The organization of this paper is as follows.
In Sec.\,\ref{sec:seesaw}, we review the seesaw mechanism and the two-zero minor structure realized in the minimal $\Umt$ model.
In Sec.\,\ref{sec:Radiative Correction on Two-Zero}, we
derive the one-loop corrections to the two-zero minor structure in the $\Umt$-symmetric model.
In Sec.\,\ref{sec:analysis}, we perform a fitting to the experimental data using the one-loop corrected neutrino mass matrix.
The final section is devoted to conclusions.

\section{Seesaw Mechanism and Two-Zero Minors}
\label{sec:seesaw}
Before discussing radiative corrections, we review the seesaw mechanism
with the two-zero minor structure at tree level.  We then discuss
the correlations among the neutrino parameters implied by the two-zero
minor structure, as well as how these correlations are modified when the
structure is slightly broken.
\subsection{Constraints on Two-Zero Minor Structure}
Let us begin with the seesaw mechanism at tree level. 
The resulting Lagrangian is given by:
\begin{align}
\label{eq:L_lepto}
&\calL
=
L^\dagger_i i\bar{\sigma}^\mu D_\mu L_i
+\bar{N}^\dagger_\alpha i\bar{\sigma}^\mu \partial_\mu \bar{N}_\alpha
+ D^\mu H^\dagger D_\mu H
-\lambda_H |H|^4
\notag\\
&\phantom{\calL=}
-\frac{1}{2} M_{\alpha\beta}\bar{N}_\alpha \bar{N}_\beta 
-\kappa_{i\alpha}\, (H\!\cdot\! L_i)\, \bar{N}_\alpha 
-y_{ij} H^\dagger \bar{E}_i L_j
+ \mathrm{h.c.}
\ ,
\\
&D_\mu L_i
=
\qty(
\partial_\mu
-\frac{i}{2} g \sigma^a W_\mu^a
+\frac{i}{2} g' B_\mu
) L_i \ ,
\\
&D_\mu H
=
\qty(
\partial_\mu
-\frac{i}{2} g \sigma^a W_\mu^a
-\frac{i}{2} g' B_\mu
) H \ ,
\\
&D_\mu \bar{E}_i = \qty(\partial_\mu - ig'B_\mu)\bar{E}_i\ , 
\\
&H\!\cdot\! L_i
=
H^+ e_{Li} - H^0 \nu_{Li} \ .
\end{align}
Here, $H$ denotes the Standard Model (SM) Higgs doublet, while
$L_i=(\nu_{Li},e_{Li})$, $\bar E_i$ $(i=1,2,3)$, and
$\bar N_\alpha$ $(\alpha=1,2,3)$ denote the lepton doublets, charged
lepton singlets, and right-handed neutrinos, respectively.  All fermions
are taken to be left-handed Weyl fields; accordingly, the bar on a fermion
field does not denote Dirac conjugation, but is simply part of the field
name.

The $\mathrm{SU}(2)_L$ and $\mathrm{U}(1)_Y$ gauge fields are denoted by
$W_\mu^a$ and $B_\mu$, respectively, where $a=1,2,3$ is the adjoint index
of $\mathrm{SU}(2)_L$.  The parameters $\lambda_H$, $\kappa_{i\alpha}$, $y_{ij}$, 
$g$, and $g'$ denote the Higgs quartic coupling, the neutrino Dirac Yukawa
couplings, the charged-lepton Yukawa
couplings, and the $\mathrm{SU}(2)_L$ and $\mathrm{U}(1)_Y$ gauge
couplings, respectively.  The right-handed neutrino Majorana masses are
encoded in a complex symmetric matrix $M_{\alpha\beta}$.

Unless otherwise stated, repeated indices are implicitly summed over
throughout this paper.  We work in the flavor basis in which the
charged-lepton Yukawa matrix $y_{ij}$ is real, positive, and diagonal.  In this
basis, we use the generation indices $i=1,2,3$ of $L_i$ and $\bar E_i$
interchangeably with the flavor labels $i=e,\mu,\tau$.

We work in the electroweak symmetric phase and integrate out the heavy right-handed neutrinos. 
The resultant dimension-five Weinberg operator~\cite{Weinberg:1979sa} is given by
\begin{align}
\mathcal{L}_{\rm eff}
\supset
-\frac{1}{2}\,C^{(0)}_{ij}\,
(H\!\cdot\! L_i)(H\!\cdot\! L_j)
+\mathrm{h.c.}\ ,
\end{align}
where $C_{ij}^{(0)}$ denotes the Wilson coefficient matrix.
In components, it is given by
\begin{align}
\label{eq:C0}
C^{(0)}_{ij}
=
-\kappa_{i\alpha}(M^{-1})_{\alpha\beta}\kappa_{j\beta}\ .
\end{align}
Once the electroweak symmetry is broken, the neutrino mass matrix is given by
\begin{align}
(M_\nu)_{ij} =  C_{ij}^{(0)}v_\mathrm{EW}^2\  .
\end{align}
Here, the vacuum expectation value (VEV) of the Higgs doublet is taken to be
$H=(0,v_\mathrm{EW})$ with $v_{\mathrm{EW}}\simeq 174$\,GeV.

In this work, we are interested in the two-zero minor structure of the Wilson coefficient matrix $C$ in the form 
\begin{align}
\label{eq:minors}
C^{-1}
=
\mqty(
*&*&*\\
*&0&*\\
*&*&0
)\ ,
\end{align}
where the indices $2$ and $3$ correspond to the $\mu$- and $\tau$-flavor components, respectively.
The two-zero minor structure is an important outcome 
of the minimal $\Umt$-symmetric model.

The neutrino mass matrix $M_\nu$ 
is a complex symmetric matrix which 
can be diagonalized using the Takagi decomposition,
\begin{align}
M_{\nu} 
= U^{*} \, \hat{m}_{\nu} \, 
U^{\dagger}\ , 
\quad \hat{m}_{\nu} = \mathrm{diag}(m_1, m_2, m_3)\ ,
\end{align}
where $m_{1,2,3}$ are taken real positive.
In the following analysis, we focus 
on the normal ordering (NO) case,
\begin{align}
0   \le m_{1}<m_{2}<m_{3}\ .
\end{align}
Since the charged-lepton sector is 
in the mass basis, 
the unitary matrix $U$  is identified with the PMNS matrix, 
\begin{align}
U
&= 
\begin{pmatrix}
U_{e1}&U_{e2}&U_{e3}\\
U_{\mu 1}&U_{\mu 2}&U_{\mu 3}\\
U_{\tau 1}&U_{\tau 2}&U_{\tau 3}
\end{pmatrix} \nonumber \\
&= 
\begin{pmatrix}
c_{12}c_{13} & s_{12} c_{13} & s_{13} e^{-i \delta_{\mathrm{CP}}} \\
- s_{12} c_{23} - c_{12} s_{13} s_{23} e^{i \delta_{\mathrm{CP}}} 
& c_{12} c_{23} - s_{12} s_{13} s_{23} e^{i \delta_{\mathrm{CP}}} 
& c_{13} s_{23} \\
s_{12} s_{23} - c_{12} s_{13} c_{23} e^{i \delta_{\mathrm{CP}}} 
& - c_{12} s_{23} - s_{12} s_{13} c_{23} e^{i \delta_{\mathrm{CP}}} 
& c_{13} c_{23}
\end{pmatrix}
\begin{pmatrix}
e^{i \eta_1} & 0 & 0 \\
0 & e^{i \eta_2} & 0 \\
0 & 0 & 1
\end{pmatrix}\ ,
\end{align}
where $c_{ij} := \mathrm{cos}\, \theta_{ij}$ and $s_{ij} := \mathrm{sin}\, \theta_{ij}$.
The phases $\delta_{\mathrm{CP}}$ and  $\eta_{1,2}$ are the Dirac and Majorana phases, respectively.

\begin{figure}[tbp]
    \centering    \includegraphics[width=0.5\linewidth]{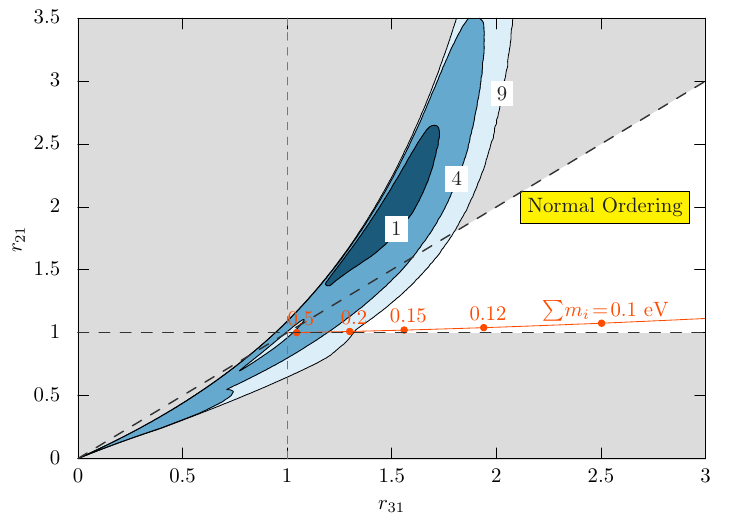}
    \caption{
Allowed region in the $(r_{31},r_{21})$ plane obtained from the
chi-square data for the neutrino mixing parameters and the Dirac CP phase
in the NuFIT~6.1 analysis (IC24 with SK-atm), assuming normal neutrino
mass ordering.  The contours correspond to
$\mathit{\Delta}\chi^2_{\mathrm{mix}}=1$, $4$, and $9$, respectively, where
$\mathit{\Delta}\chi^2_{\mathrm{mix}}$ is defined in
Eq.\,\eqref{eq:chisquare mixing}.  The observed values of
$\mathit{\Delta} m_{21}^2$ and $\mathit{\Delta} m_{31}^2$ are not used in drawing the
contours.  The red solid line shows the relation in
Eq.\,\eqref{eq:r21 prediction}, obtained by imposing the observed ratio
$\mathit{\Delta} m_{21}^2/\mathit{\Delta} m_{31}^2$.  The numbers attached to the points on
the red solid line denote the corresponding values of the total neutrino
mass $\sum_i m_i$.
}
\label{fig:r31-r21}
\end{figure}

The  two zeros of
$M_\nu^{-1}$ 
in the $(\mu,\mu)$ and $(\tau,\tau)$ entries 
lead to 
the following relation
between the neutrino mass ratios and  the mixing angles and CP phases
\,\cite{Verma:2011kz,Liao:2013saa,Asai:2017ryy},

\begin{align}
&r_{21}:=\frac{m_2}{m_1}=
    \frac{
c_{12}
\left[
-c_{12}\cos2\theta_{23}
+
e^{i\delta_\mathrm{CP}}s_{12}s_{13}\sin2\theta_{23}
\right]
}{
s_{12}
\left[
\cos2\theta_{23}s_{12}
+
e^{i\delta_\mathrm{CP}}c_{12}s_{13}\sin2\theta_{23}
\right]
}
    \times e^{-2i(\eta_1-\eta_2)}
    \ ,
    \label{eq:r21}
    \\
&r_{31}:=\frac{m_3}{m_1}
= 
\frac{
c_{12} c_{13}^{2}
\left(
c_{12}\cos 2\theta_{23}
-
2 e^{i \delta_\mathrm{CP}}
c_{23}s_{12}s_{13}s_{23}
\right)
}{
e^{i \delta_\mathrm{CP}}s_{13}
\left[
-
e^{i \delta_\mathrm{CP}}
\cos 2\theta_{12}
\cos 2\theta_{23}
s_{13}
+
2c_{12}c_{23}s_{12}
\left(
1+e^{2i\delta_\mathrm{CP}}s_{13}^{2}
\right)
s_{23}
\right]
} \times e^{-2 i \eta_1}
\ .
\label{eq:r31}
\end{align}
Note that $r_{21}$ and $r_{31}$ are real and positive, which 
fixes the value of 
the CP phases $\eta_1$ and $\eta_2$ for a given $\delta_\mathrm{CP}$. 
These relations impose strong correlations among the mixing angles, the Dirac CP phase, and the mass ratios.

In Fig.\,\ref{fig:r31-r21}, we show the allowed region in the
$(r_{31},r_{21})$ plane, obtained using the chi-square data for the
neutrino mixing parameters and the Dirac CP phase from the NuFIT~6.1
analysis (IC24 with SK-atm).  
The details of our statistical treatment of
the chi-square data are summarized in the Appendix~\ref{app:chi2}.  
In the figure, we use
the data for the normal neutrino mass ordering, for which the consistent
region corresponds to $r_{31}>r_{21}>1$.  The contours show
$\mathit{\Delta}\chi^2_{\mathrm{mix}}=1$, $4$, and $9$, respectively, where
$\mathit{\Delta}\chi^2_{\mathrm{mix}}$ is defined in Eq.\,\eqref{eq:chisquare mixing}.
Notice that the observed values of $\mathit{\Delta} m_{21}^2$ and
$\mathit{\Delta} m_{31}^2$ are not used in drawing these contours.

The figure shows that $r_{31}$ remains of order unity even for
$\mathit{\Delta}\chi^2_{\mathrm{mix}}=9$.  This gives a very strong constraint on
$r_{21}$.  From
\begin{align}
\frac{\mathit{\Delta} m_{21}^2}{\mathit{\Delta} m_{31}^2}
=
\frac{r_{21}^2-1}{r_{31}^2-1}
\end{align}
and the observed ratio
$\mathit{\Delta} m_{21}^2/\mathit{\Delta} m_{31}^2\sim 3\times 10^{-2}$, we obtain, for $r_{31}=\order{1}$,
\begin{align}
\label{eq:r21 prediction}
r_{21}
&=
\sqrt{
1+(r_{31}^2-1)
\frac{\mathit{\Delta} m_{21}^2}{\mathit{\Delta} m_{31}^2}
}
\nonumber\\
&\simeq
1+
\frac{r_{31}^2-1}{2}
\frac{\mathit{\Delta} m_{21}^2}{\mathit{\Delta} m_{31}^2}\ .
\end{align}
Thus, once the observed values of $\mathit{\Delta} m_{21}^2$ and
$\mathit{\Delta} m_{31}^2$ are imposed, $r_{21}$ is required to be very close to
unity.  The horizontal red solid line in the figure shows
Eq.\,\eqref{eq:r21 prediction} as a function of $r_{31}$.  Therefore, the
parameter region allowed by both the mass-squared differences and the
mixing parameters, including the Dirac CP phase, is given only by the
intersection of the $\mathit{\Delta}\chi^2_{\mathrm{mix}}$ allowed region with the
red solid line.

It is also useful to note that, for a given value of $r_{31}$, the total
neutrino mass is predicted.  Indeed, for normal ordering we obtain
\begin{align}
\sum_i m_i= \sqrt{\frac{\Delta m_{31}^2}{r_{31}^2-1}}
\qty(1+r_{31}+\sqrt{1+(r_{31}^2-1)\frac{\Delta m_{21}^2}{\Delta m_{31}^2}})\ .
\end{align}
The numbers
attached to the points on the red solid line in the figure denote the
corresponding values of the total neutrino mass determined by this
relation.  From the figure, we can see that the total neutrino mass
becomes large near the region where the black solid line intersects the
$\mathit{\Delta}\chi^2_{\mathrm{mix}}$ contours,
\begin{align}
    \sum_{i}m_i \gtrsim 0.2\,\mathrm{eV}\ ,
\end{align}
which reproduces the analysis in Ref.\,\cite{Ibe:2016dir}.

\subsection{Impact of 
Deviations from the Two-Zero Minor Structure}
Before discussing the details of the radiative corrections,
let us discuss the impact of deviations from the
two-zero minor structure.
For this purpose, we define
\begin{align}
\label{eq:epsilon mumu}
\epsilon_{\mu\mu} &:=     e^{-2i\eta_1}m_1 \times (M_{\nu}^{-1})_{\mu\mu}\ , \\
\label{eq:epsilon tautau}
\epsilon_{\tau\tau} &:=     e^{-2i\eta_1}m_1\times (M_{\nu}^{-1})_{\tau\tau}\ , \ 
\end{align}
which parameterize the deviation from the two-zero minor structure.
Here, $\epsilon_{\mu\mu}$ 
and $\epsilon_{\tau\tau}$
are small complex-valued parameters.
At tree level,
$\epsilon_{\mu\mu}=\epsilon_{\tau\tau}=0$.
As we will see later, however,
radiative corrections lead to
$\epsilon_{\mu\mu}\neq 0$
and
$\epsilon_{\tau\tau}\neq 0$,
which can be of $\order{0.1}$ depending on the model parameters.

\begin{figure}[tbp]
    \centering    \includegraphics[width=0.5\linewidth]{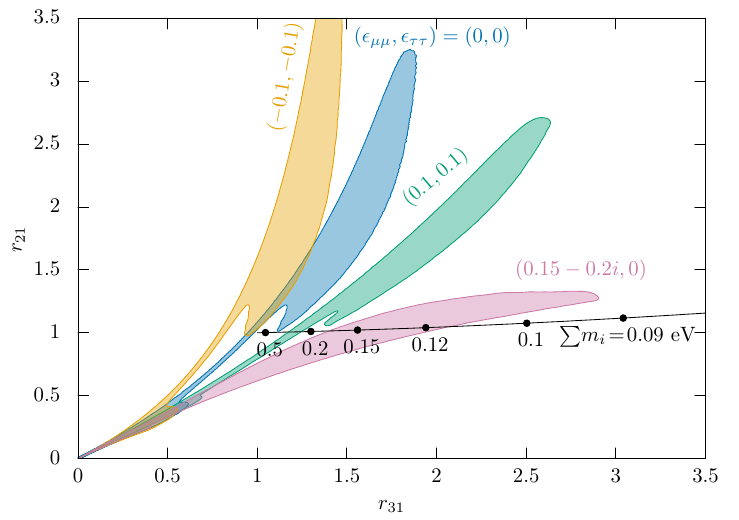}
    \caption{
    Allowed regions in the
$(r_{31}^{(\epsilon)},r_{21}^{(\epsilon)})$ plane for fixed values of
$(\epsilon_{\mu\mu},\epsilon_{\tau\tau})$, shown in the plot.  For clarity, only
    the contour corresponding to $\mathit{\Delta}\chi^2_{\mathrm{mix}}=3$ is shown.
Since we use
the data for the normal ordering, the consistent region corresponds to $r_{31}>r_{21}>1$.
    The case $(\epsilon_{\mu\mu},\epsilon_{\tau\tau})=(0,0)$ corresponds
    to the exact two-zero minor structure and reproduces the result in
    Fig.\,\ref{fig:r31-r21}.  The black solid line is the same as that in
    Fig.\,\ref{fig:r31-r21}, which is obtained by imposing the observed
    ratio $\mathit{\Delta} m_{21}^2/\Delta m_{31}^2$.
    }
    \label{fig:r31-r21-epsilon}
\end{figure}

In the presence of the $\epsilon$'s,
the mass ratios $r_{21}$
and $r_{31}$ are modified as
\begin{align}
r_{21}^{(\epsilon)}
&=
\frac{
c_{12}
\left[
-c_{12}\cos2\theta_{23}
+
e^{i\delta_\mathrm{CP}}s_{12}s_{13}\sin2\theta_{23}
\right]
}{
s_{12}
\left[
\cos2\theta_{23}s_{12}
+
e^{i\delta_\mathrm{CP}}c_{12}s_{13}\sin2\theta_{23}
\right]
-
c_{23}^2\,\epsilon_{\mu\mu}
+
s_{23}^2\,\epsilon_{\tau\tau}
}
\times e^{-2 i(\eta_1-\eta_2)}
\ , \label{eq:r21_ep}\\
r_{31}^{(\epsilon)} &=
\frac{
c_{12} c_{13}^{2}
\left(
c_{12}\cos 2\theta_{23}
-
2 e^{i \delta_\mathrm{CP}}
c_{23}s_{12}s_{13}s_{23}
\right)
}{
e^{i \delta_\mathrm{CP}}s_{13}
\left[
-
e^{i \delta_\mathrm{CP}}
\cos 2\theta_{12}
\cos 2\theta_{23}
s_{13}
+
2c_{12}c_{23}s_{12}
\left(
1+e^{2i\delta_\mathrm{CP}}s_{13}^{2}
\right)
s_{23}
\right]
-K_{31}
} \times e^{-2 i \eta_1}
\ , \label{eq:r31_ep}\\
K_{31}&= \left(
e^{i \delta_\mathrm{CP}}c_{23}s_{12}s_{13}
+
c_{12}s_{23}
\right)^{2}
\epsilon_{\mu\mu}
-
\left(
c_{12}c_{23}
-
e^{i \delta_\mathrm{CP}}s_{12}s_{13}s_{23}
\right)^{2}
\epsilon_{\tau\tau}
\ .
\end{align}

To illustrate the impact of nonzero $\epsilon_{\mu\mu}$ and
$\epsilon_{\tau\tau}$, we repeat the analysis in the
$(r_{31}^{(\epsilon)},r_{21}^{(\epsilon)})$ plane for several fixed
values of $(\epsilon_{\mu\mu},\epsilon_{\tau\tau})$.  The case
$(\epsilon_{\mu\mu},\epsilon_{\tau\tau})=(0,0)$ corresponds to the exact
two-zero minor structure and reproduces the result shown in
Fig.\,\ref{fig:r31-r21}.  In order to avoid overcrowding the plots, we
show only the contour corresponding to
$\mathit{\Delta}\chi^2_{\mathrm{mix}}=3$ in the plot. The black solid line shows the relation in Eq.\,\eqref{eq:r21 prediction} as in the case of Fig.\,\ref{fig:r31-r21}.

The figure shows that the allowed region determined from the neutrino
mixing parameters and the Dirac CP phase can be considerably modified by
introducing $\epsilon_{\mu\mu}$ and $\epsilon_{\tau\tau}$ of
$\order{0.1}$.  In particular, depending on the phases of
$\epsilon_{\mu\mu}$ and $\epsilon_{\tau\tau}$, the intersection between
the black solid line and the allowed contour can be shifted toward larger
values of $r_{31}$.

Since a larger value of $r_{31}$ corresponds to a smaller total neutrino
mass along the black solid line, this shift implies that the lower bound
on the total neutrino mass obtained in the previous section can be
significantly relaxed.  In the next section, we show that radiative
corrections can indeed generate deviations of this size.

\section{Radiative Corrections to the Two-Zero Minor Structure}
\label{sec:Radiative Correction on Two-Zero}

So far, we have discussed the two-zero minor structure without specifying a concrete ultraviolet (UV) theory that realizes it. 
However, in order for the analysis of radiative corrections to the minor structure to be physically meaningful, the tree-level zero-minor structure should be realized, or protected, by an underlying symmetry.
In other words, zero structures that are not protected by an underlying symmetry are not physical by themselves, but rather depend on the choice of renormalization conditions.

\subsection{Minimal \texorpdfstring{$\boldsymbol{\Umt}$}{U(1)LmuLtau} Model}

Motivated by the above observation, we now specify a class of UV models in which the two-zero minor structure is enforced by a symmetry.
As a concrete example, we consider a model based on the $\Umt$ gauge symmetry or one of its discrete subgroups.
The charge assignment of the minimal model is given in Table~\ref{tab:mutau_charge}, where the symmetry is spontaneously broken by the vacuum expectation value of a single $\Umt$-charged scalar field $\phi$~\cite{Asai:2017ryy,Asai:2018ocx,Asai:2020qax}.

In this setup, both the charged-lepton Yukawa coupling
matrix $y$
and the neutrino Dirac Yukawa coupling matrix $\kappa$ are diagonal as a consequence of the symmetry. 
The right-handed neutrino mass matrix,
on the other hand, has vanishing $(\mu,\mu)$ and $(\tau,\tau)$ components. 
Explicitly, they have the form,
\begin{align}
y =
\mqty(
y_e & 0 & 0 \\
0 & y_\mu & 0 \\
0 & 0 & y_\tau
)\ ,\qquad
\kappa =
\mqty(
\kappa_e & 0 & 0 \\
0 & \kappa_\mu & 0 \\
0 & 0 & \kappa_\tau
)\ ,
\qquad
M =
\mqty(
M_{ee} & \lambda_{e\mu} \expval{\phi} & \lambda_{e\tau} \expval{\phi^*} \\
\lambda_{e\mu} \expval{\phi} & 0 & M_{\mu\tau} \\
\lambda_{e\tau} \expval{\phi^*} & M_{\mu\tau} & 0
)\ .
\label{eq:mu tau param}
\end{align}
Here, $\lambda_{e\mu}$ and $\lambda_{e\tau}$ are Yukawa couplings associated with the $\Umt$-breaking field $\phi$.
The right-handed neutrino mass terms allowed before $\Umt$ breaking are $M_{ee}$ and $M_{\mu\tau}$, whereas the remaining nonzero components are generated after the spontaneous breaking of $\Umt$.
As a result, the Wilson coefficient matrix $C^{(0)} = -\kappa M^{-1}\kappa^T$ possesses the two-zero minor structure shown in Eq.\,\eqref{eq:minors} at tree level.

Note that the couplings $\kappa_i$ and $y_i$ can be made real and positive by utilizing the phase freedom of the lepton doublets and singlets. 
Furthermore, using appropriate phase redefinitions of the fields, one can also choose all entries of the right-handed neutrino mass matrix except for $M_{ee}$ to be real and positive.
Consequently, the minimal
$\Umt$ model contains a single physical parameter responsible for CP violation. 
Therefore, the model has eight real-valued parameters 
associated with the right-handed neutrinos.

\begin{table}[t!]
\begin{center}
\renewcommand{\arraystretch}{1.5}
\caption{
The $\Umt$ charge assignments in the minimal $\Umt$ gauge theory. 
All other SM fields are neutral under this symmetry.
We may consider an appropriate discrete subgroup of $\Umt$ gauge symmetry to achieve the two-zero minor structure.
}
\begin{tabular}{ccc} \hline
   &Field & $\Umt$ \\ \hline
   \multirow{3}{*}{Leptons} 
    & $L_{e}$ \,\,\, $\ebar_{e}$ \,\,\, $\Nbar_{e}$ & $0$ \\
   & $L_{\mu}$ \,\,\, $\ebar_{\tau}$ \,\,\, $\Nbar_{\tau}$ & $+1$ \\
    & $L_{\tau}$ \,\,\, $\ebar_{\mu}$ \,\,\, $\Nbar_{\mu}$ & $-1$ \\ \hline
   \multirow{1}{*}{Scalar} & $\phi$ & $+1$ \\ \hline
\end{tabular}
\label{tab:mutau_charge}
\end{center}
\end{table}

\subsection{Universal One-Loop Corrections}
We now discuss radiative corrections to the two-zero minor structure in the $\Umt$ model.
To perform the one-loop analysis, it is useful to diagonalize the tree-level right-handed neutrino mass matrix by the Takagi decomposition,
\begin{align}
\bar{N}_\alpha = V_{\alpha I}\hat{\bar{N}}_I \ .
\end{align}
Here, $\alpha=1,2,3$ denotes the flavor index in the original basis, while $I=1,2,3$ labels the mass eigenstates.
The unitary matrix $V$ satisfies
\begin{align}
(V^T M V)_{IJ} = V_{\alpha I} M_{\alpha\beta} V_{\beta J}
= \hat{M}_I \delta_{IJ} \ ,
\end{align}
with
\begin{align}
\hat{M}_I > 0 \ .
\end{align}
In this basis, the Dirac Yukawa couplings are in general non-diagonal,
\begin{align}
\hat{\kappa}_{iI} = \kappa_{i\alpha} V_{\alpha I} \ ,
\end{align}
and the tree-level Wilson coefficient for the  neutrino mass operator is given by
\begin{align}
    C^{(0)}_{ij} = -\kappa_{i\alpha}(M^{-1})_{\alpha\beta}\kappa_{j\beta}
    = - \frac{\hat{\kappa}_{iI}\hat{\kappa}_{jI}}{\hat{M}_I}\ .
\end{align}

The one-loop 1PI effective Lagrangian relevant to the seesaw mechanism is given by
\begin{align}
\label{eq:Leff}
\calL_\mathrm{1PI} =& 
L^\dagger_i 
i\bar{\sigma}^\mu D_\mu\qty(\delta_{ij}- \Sigma^L_{ij}) L_j
+\bar{N}^\dagger_\alpha 
i\bar{\sigma}^\mu \partial_\mu
\qty( \delta_{\alpha\beta}-\Sigma^N_{\alpha\beta})
\bar{N}_\beta
+ D^\mu H^\dagger D_\mu H \qty(1-\Sigma^H)
\cr
&- \frac{1}{2} 
\hat{\bar{N}}_I\qty( \hat{M}_{I}\delta_{IJ}+\Omega_{IJ} )\hat{\bar{N}}_J
- 
\qty(\hat{\kappa}_{iI}+{\Gamma}_{HL_i\hat{N}_I}) (H\!\cdot\!L_i) \hat{\bar{N}}_I 
-H^\dagger \bar{E}_i y_i\qty(\delta_{ij} + \Gamma_{H\bar{E}_iL_j})L_j\cr
&-
\frac{1}{2}\Gamma_{HL_iHL_j}(H\!\cdot\!L_i)(H\!\cdot\!L_j)
+\mathrm{h.c.}
\end{align}
The quantities $\Sigma$'s, $\Omega$, $\Gamma_{HL\hat{N}}$, $\Gamma_{H\bar{E}_iL_j}$, and $\Gamma_{HLHL}$ denote the renormalized 1PI self-energies, the $\mathit{\Delta}L=2$ two-point function, the three-point functions, and the four-point function, respectively.
The four-point function is induced by the box diagrams,
which corresponds to the finite correction to the neutrino mass $\delta M_L$ in Refs.\,\cite{Grimus:2002nk,AristizabalSierra:2011mn}.

An important point is that, in order to cancel the UV divergences, one does not need generic counterterms and those that preserve $\Umt$ symmetry are sufficient. This property ensures that the theory retains its predictive power for radiative corrections that break the two-zero minor structure.

In our analysis, we focus on the corrections induced by the interactions appearing in Eq.\,\eqref{eq:L_lepto}.
These corrections are universal for 
the type-I seesaw models
in the sense that they are present independently of the details of the UV model.
In addition to these universal effects, the $\Umt$ gauge interaction and the interactions in the $\Umt$-breaking sector can also induce radiative corrections (see e.g., Ref.\,\cite{Ibarra:2026max}).
These radiative corrections generically modify the two-zero minor structure of the neutrino mass matrix.
Such effects are, however, model dependent, that is, they depend not only on the coupling constants in the 
$\Umt$-breaking sector, but also on whether the symmetry realizing the minor structure is the gauged $\Umt$ symmetry or one of its discrete subgroups.
We therefore do not include them in the following analysis, assuming that the relevant couplings in these sectors are sufficiently small.

Under this assumption, the one-loop corrections relevant to the minor structure are classified as follows:
\begin{itemize}
\item box diagrams induced by the $\mathrm{SU}(2)_L$ and $\mathrm{U}(1)_Y$ gauge interactions, with right-handed neutrinos running in the loop;
\item box diagrams involving the Higgs doublet and right-handed neutrinos;
\item $\mathcal{O}(\kappa^2)$ corrections to the kinetic terms of the lepton doublets and the Higgs doublet;
\item $\mathcal{O}(\kappa^2)$ corrections to the charged-lepton Yukawa couplings.
\end{itemize}
As stated earlier, we work in the electroweak symmetric phase, 
which simplifies the 
analysis compared with previous one-loop analyses in Refs.\,\cite{Grimus:1989pu,
Pilaftsis:1991ug,
Kniehl:1996bd,
Grimus:2002nk,AristizabalSierra:2011mn}.
In this phase, the interactions in Eq.\,\eqref{eq:L_lepto} do not induce three-point vertex corrections to $\hat{\kappa}$.
We also note that they do not generate additional $\mathit{\Delta} L=2$ corrections at the one-loop order.

With these simplifications, the  one-loop contributions to the Wilson coefficient are given by
\begin{align}
\label{eq:Cij}
    C^{(1)}_{ij}
  =
  \frac{1}{2}\hat{\Sigma}^L_{ik}C_{kj}^{(0)}
  +  \frac{1}{2}C_{ik}^{(0)}\hat{\Sigma}^L_{kj}
  +
C_{ij}^{(0)}\Sigma^H
    +\Gamma_{HL_iHL_j}\ .
\end{align}
Here, $\hat{\Sigma}^L$ denotes the correction to the lepton-doublet kinetic term in the basis where the charged-lepton Yukawa coupling is diagonal at the one-loop level 
(see Appendix~\ref{app:diagonal lepton}).

In our analysis, we match the Wilson coefficient obtained at one-loop order in the UV theory onto the corresponding Wilson coefficient in the effective field theory (EFT), where the right-handed neutrinos have been integrated out.
Therefore, radiative corrections that are common to both UV theory and EFT need not be included explicitly.
One caveat is that, since we perform the calculation in the electroweak symmetric phase, infrared (IR) divergences appear in radiative corrections both in the UV theory and in the EFT.
These IR divergences cancel in the matching procedure.

Finally, radiative corrections induced by the charged-lepton Yukawa couplings do not appear in the matching condition considered here at the one-loop level. On the other hand, flavor-dependent radiative corrections to the charged-lepton Yukawa coupling are relevant for correctly defining the lepton-doublet basis in which the charged-lepton Yukawa matrix remains diagonal at the one-loop level. This point is discussed in Appendix~\ref{app:diagonal lepton}.

\subsection{Evaluation of the One-Loop Corrections}
\subsubsection*{SU(2)\texorpdfstring{$_L$}{L} and U(1)\texorpdfstring{$_Y$}{Y} Gauge Boson Box Contributions}
\begin{figure}[tbp]
\centering
\includegraphics[width=0.7\linewidth]{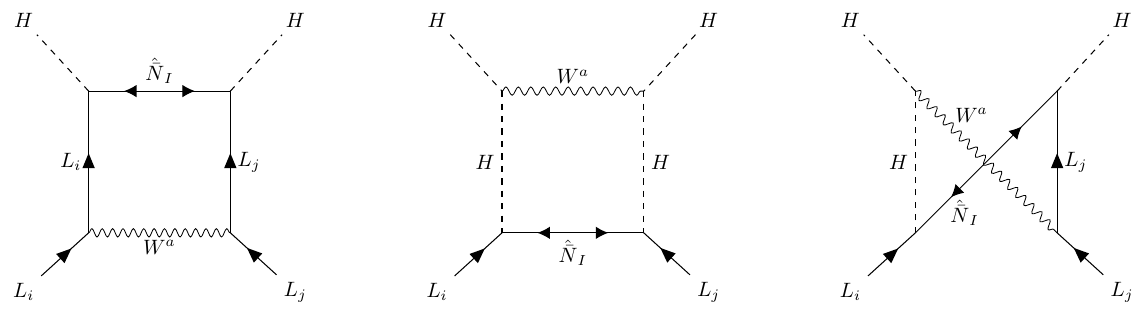}
\caption{The $\mathrm{SU}(2)_L$ box diagram contributions to the Weinberg operator for neutrino masses. 
There are also analogous $\mathrm{U}(1)_Y$ box diagram contributions.}
\label{fig:Z-box}
\end{figure}

Let us first consider the gauge-boson box diagram contributions shown in Fig.\,\ref{fig:Z-box}.
We adopt a $\xi$-gauge for the gauge interactions since we are in the electroweak symmetric phase.

Concretely, the gauge-field propagator is given by
\begin{align}
    iD_{\mu\nu}
    =\frac{-i}{p^2-m_\mathrm{IR}^2+i\varepsilon}\qty(g_{\mu\nu}-(1-\xi)\frac{p_\mu p_\nu}{p^2}) \ ,
\end{align}
where we have introduced an IR cutoff scale $m_\mathrm{IR}$ to regularize the IR divergence.
Similarly, we put the IR cutoff in the propagator of the Higgs and lepton doublets.

Combining the contributions of the three diagrams, we obtain the $\mathrm{SU}(2)_L$ gauge-interaction contribution,
\begin{align}
      \Gamma_{HLHL}^{(W)} 
      = 
      -\frac{3g^2}{64\pi^2}
\hat{\kappa}\hat{M}^{-1}
\qty(\frac{3}{2}-\ln\frac{\hat{M}^2}{m_\mathrm{IR}^2})\hat{\kappa}^T \ .
\end{align}
Here, we have used the $3\times 3$ matrix notation and introduced $\hat{M}=\mathrm{diag}(\hat{M}_1,\hat{M}_2,\hat{M_3})$.
We omit the unit matrix.
This contribution is UV finite but IR divergent.
Similarly, the $\mathrm{U}(1)_Y$ contribution is given by
\begin{align}
      \Gamma_{HLHL}^{(B)}= 
         -\frac{3g^{\prime2}}{64\pi^2}
\hat{\kappa}\hat{M}^{-1}
\qty(\frac{3}{2}-\ln\frac{\hat{M}^2}{m_\mathrm{IR}^2})\hat{\kappa}^T  \ .
\end{align}
Combining the $\mathrm{SU}(2)_L$ and $\mathrm{U}(1)_Y$ contributions, we obtain
\begin{align}
\Gamma_{HLHL}^{(W, B)}  =  -\frac{3(g^2+g^{\prime2})^2}{64\pi^2}
\hat{\kappa}\hat{M}^{-1}
\qty(\frac{3}{2}-\ln\frac{\hat{M}^2}{m_\mathrm{IR}^2})\hat{\kappa}^T\ .
\end{align}

To match the above result onto the Wilson coefficient in the EFT, we consider the effective operator
\begin{align}
    \calL =-\frac{1}{2} C^{(\mathrm{EFT},0)}_{ij} (H\!\cdot\!L_i) (H\!\cdot\!L_j) \ ,
\end{align}
where, at tree level,
\begin{align}
    C^{(\mathrm{EFT},0)}_{ij} = C^{(0)}_{ij}= -\frac{\hat{\kappa}_{iI}\hat{\kappa}_{jI}}{\hat{M}_I} \ .
\end{align}
Repeating the above analysis in the EFT, we obtain
\begin{align}
C^{(\mathrm{EFT},1)}|_\mathrm{gauge} = -  \frac{3(g^2+g^{\prime 2})}{64\pi^2 }    
C^{(0)}
\qty(\frac{1}
    {\bar{\epsilon}}+\ln\frac{\mu^2}{m_\mathrm{IR}^2}-\frac{7}{6})\ .
\end{align}

\subsubsection*{Higgs Boson Box Contribution}
\begin{figure}
    \centering    
    \includegraphics[width=0.18\linewidth]{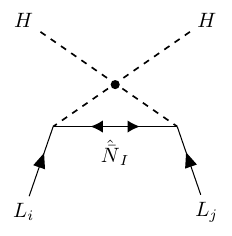}
    \caption{Higgs boson contribution to the Weinberg operator for neutrino masses.}
    \label{fig:H-box}
\end{figure}
The Higgs contribution through the box diagram in Fig.\,\ref{fig:H-box} is given by
\begin{align}
\Gamma^{(H)}_{HLHL} &=-\frac{\lambda_H}{8\pi^2}\hat{\kappa}\hat{M}^{-1}
\qty(1-\ln\frac{\hat{M}^2}{m_\mathrm{IR}^2})\hat{\kappa}^T\ .
\end{align}
In the EFT, the corresponding diagram gives
\begin{align}
    C^{(\mathrm{EFT},1)}|_\mathrm{Higgs} = -
\frac{\lambda_H}{8\pi^2 } C^{(0)} \qty(\frac{1}{\bar{\epsilon}}+\ln\frac{\mu^2}{m_\mathrm{IR}^2})\ .
\end{align}

\subsubsection*{Self-Energy Contributions}
\begin{figure}[tbp]
\centering
\includegraphics[width=0.5\linewidth]{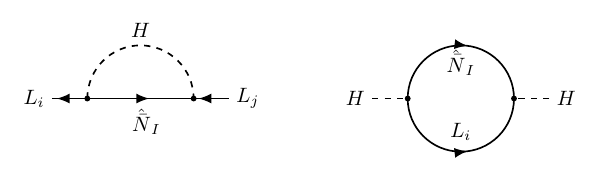}
\caption{The self-energy contributions.}
\label{fig:wavefunction}
\end{figure}

The two-point self-energy contributions $\Sigma^L$ and $\Sigma^H$ defined in Eq.\,\eqref{eq:Leff} are generated by the diagrams in Fig.\,\ref{fig:wavefunction}.
At vanishing external momentum, the self-energy of the charged leptons is given by
\begin{align}
\label{eq:SigmaLbare}
 \Sigma^L_{ij} = 
   - \frac{1}{2}\qty(\delta_{L_i} + \delta_{L_j})
     -\frac{\hat{\kappa}_{iI}^* \hat{\kappa}_{jI}}{32\pi^2} \qty(\frac{1}{\bar{\epsilon}}+ \ln\frac{\mu^2}{\hat{M}_I^2}+\frac{3}{2})\ .
\end{align}
Here, we have also included the wave-function renormalization factor,
\begin{align}
\label{eq:bareL} 
    L_{iB} = L_{i}\qty(1+\frac{1}{2}\delta_{L_i})\ ,
\end{align}
where $L_{iB}$ denotes the bare lepton doublet.
Due to the $\Umt$ symmetry, the flavor-diagonal wave function renormalization is enough to cancel the UV divergence. 
In the $\MSB$ scheme, we obtain
\begin{align}
\label{eq:SigmaL}
 \Sigma^L = 
    -\frac{1}{32\pi^2} \hat{\kappa}^*\qty(\ln\frac{\mu^2}{\hat{M}^2}+\frac{3}{2})\hat{\kappa}^
    T\ .
\end{align}

The self-energy of the Higgs doublet is given by
\begin{align}
    \Sigma^H =-\delta_H -\frac{1}{16\pi^2}
   \tr[\kappa^* \qty(\frac{1}{\bar{\epsilon}}+\ln\frac{\mu^2}{\hat{M}^2} + \frac{1}{2})\kappa^T]\ .
\end{align}
Here, we have introduced the wave-function renormalization factor of $H$,
\begin{align}
    H_B = \qty(1+\frac{1}{2}\delta_H)H \ ,
\end{align}
with $H_B$ being the bare Higgs doublet.
In the $\MSB$ scheme, $\Sigma^H$ reduces to
\begin{align}
\label{eq:SigmaH}
    \Sigma^H=  -\frac{1}{16\pi^2}
   \tr[\kappa^* \qty(\ln\frac{\mu^2}{\hat{M}^2} + \frac{1}{2})\kappa^T]\ .
\end{align}
In the EFT side, the corresponding $\Sigma^{L,H}$ vanish.
\subsubsection*{Correction to Charged-Lepton Yukawa Couplings}
\begin{figure}[tbp]
\centering
\includegraphics[width=0.22\linewidth]{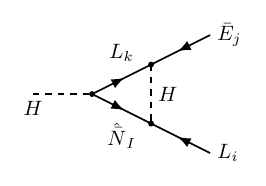}
\caption{The correction to charged-lepton Yukawa couplings.}
\label{fig:chargedleptonYukawa}
\end{figure}
Finally, let us consider the vertex correction from the Dirac neutrino Yukawa coupling.
These corrections are relevant 
for one-loop matching where the charged-lepton Yukawa couplings in the EFT are taken to be flavor diagonal.

At the one-loop level,  $\Gamma_{H\bar{E}_iL_j}$ is given by
\begin{align}
    \Gamma_{H\bar{E}_iL_j}=
    \qty(\delta_{y_i}
     + \frac{1}{2}\delta_{L_i}
     +\frac{1}{2}\delta_H)
     \delta_{ij} 
     + \frac{\hat{\kappa}_{iI}^*\hat{\kappa}_{jI}}{16\pi^2} \qty(\frac{1}{\bar{\epsilon}} + 
   \ln \frac{\mu^2}{\hat{M}_I^2}+1) \ .
\end{align}
Here, $\delta_{y_i}$ is a renormalization factor of the diagonal charged Yukawa coupling, $y_{iB}=(1+\delta_{y_i})y_{i}$.
In the $\MSB$ scheme, $\Gamma_{H\bar{E}L}$ is reduced to
\begin{align}
\label{eq:GammaHEL}
\Gamma_{H\bar{E}L} =
\frac{1}{16\pi^2} 
\hat{\kappa}^*\qty(\ln \frac{\mu^2}{\hat{M}^2}+1)\hat{\kappa}^T \ .
\end{align}
In the EFT, the corresponding $\Gamma_{H\bar{E}L}$ vanishes in the limit of vanishing external momenta in the electroweak symmetric phase.

\subsection{Matching Between UV Theory and EFT}
Let us now consider the matching between the UV theory and the EFT.
We take the EFT to be defined such that the kinetic terms of the Higgs and lepton doublets are canonical at tree level. 
We also choose a basis in which the tree-level charged-lepton Yukawa coupling is diagonal. 
For this purpose, we use fields redefined from the $\overline{\mathrm{MS}}$-normalized fields computed in the previous subsection (see Appendix~\ref{app:diagonal lepton}).

Then, putting all the contributions together, the one-loop correction to the Wilson coefficient Eq.\,\eqref{eq:Cij} in the UV theory is given by
\begin{align}
    C^{(1)} =\ &
\frac{1}{2}(\theta^L)^T C^{(0)}
+
\frac{1}{2}C^{(0)} \theta^L\cr
&-\frac{1}{16\pi^2}
\tr[\hat{\kappa}^* \qty(\ln\frac{\mu_M^2}{\hat{M}^2} + \frac{1}{2})\hat{\kappa}^T]C^{(0)}\cr
&-\frac{1}{32\pi^2}\qty( C^{(0)} \hat{\kappa}^*\qty(\ln\frac{\mu_M^2}{\hat{M}^2}+\frac{3}{2})\hat{\kappa}^T
+ \hat{\kappa}\qty(\ln\frac{\mu_M^2}{\hat{M}^2}+\frac{3}{2})\hat{\kappa}^\dagger C^{(0)})\cr
&- 
\frac{1}{64\pi^2}
\hat{\kappa} \hat{M}^{-1}
\qty(
3(g^2 + g^{\prime 2})
\qty(\frac{3}{2}-\ln\frac{\hat{M}^2}{m_\mathrm{IR}^2})+
8\lambda_H
\qty(1-\ln\frac{\hat{M}^2}{m_\mathrm{IR}^2})
)\hat{\kappa}^T \ ,
\label{eq:C1ij}
\end{align}
at the matching scale $\mu_M = \mathcal{O}(\hat{M})$.
Here, $\theta^L$ can be read off from Eqs.\,\eqref{eq:thetaL} as 
\begin{align}
\label{eq:thetaLII}
    \frac{1}{2}\theta_L\simeq 
\mqty(0&\hat{\Delta}_{e\mu}&
\hat{\Delta}_{e\tau}\\
    -\hat{\Delta}^*_{e\mu}&0&
    \hat{\Delta}_{\mu\tau}\\
    -\hat{\Delta}^*_{e\tau}&-\hat{\Delta}^{*}_{\mu\tau}&0
    )\ , 
\end{align}
with
\begin{align}
\label{eq:Deltabar}
    \hat{\Delta}
    = \frac{1}{16\pi^2} \hat{\kappa}\qty( \frac{3}{4}\ln \frac{\mu_M^2}{\hat{M}^2} + \frac{5}{8})\hat{\kappa}^\dagger \ .
\end{align}

In the EFT, on the other hand, the corresponding one-loop corrections add up to
\begin{align}
C^{(\mathrm{EFT},1)} &= 
    \delta C^{\mathrm{(match)}} 
    +  \delta C^{\mathrm{(\MSB)}}
    -  \qty(\frac{3(g^2 + g^{\prime 2})}{64\pi^2}   \qty(\frac{1}
{\bar{\epsilon}}+\ln\frac{\mu_M^2}{m_\mathrm{IR}^2}-\frac{7}{6})+
    \frac{\lambda_H}{8\pi^2} \qty(\frac{1}{\bar{\epsilon}}+\ln\frac{\mu_M^2}{m_\mathrm{IR}^2}))
    C^{(0)}\ . 
\label{eq:C1EFTij}
\end{align}
Here, we have separated the EFT counterterm into the finite matching part and the $\overline{\mathrm{MS}}$ part.
By equating Eq.\,\eqref{eq:C1ij} and Eq.\,\eqref{eq:C1EFTij}, we fix $\delta C^{\mathrm{(match)}}$ 
at the matching scale $\mu_M = \mathcal{O}(\hat{M})$.
Notice that 
the IR divergences in Eqs.\,\eqref{eq:C1ij} and \eqref{eq:C1EFTij}
cancel each other, and hence, 
$\delta C^{(\mathrm{match})}$
is IR finite.

Once $\delta C^{\mathrm{(match})}$ is fixed by the matching, 
the one-loop prediction for the neutrino mass is given by
\begin{align}
M_\nu = C^{(\MSB)}(\mu)v_\mathrm{EW}^2  + 
\cdots \ .
\label{eq:CFULL}
\end{align}
Here, 
$C^{(\MSB)}(\mu)$
is the Wilson coefficient 
in the $\MSB$ scheme
with the boundary condition at the matching scale $\mu_M$,
\begin{align}
\label{eq:CMSB}
    C^{(\MSB)}(\mu_M) =  C^{(0)} + \delta C^\mathrm{(match)}\ ,
\end{align}
evolved down to $\mu$
following the
renormalization group (RG) equation given in Ref.\,\cite{Antusch:2001ck}.
The ellipses denote the
one-loop corrections to the Wilson coefficient in the electroweak symmetry broken phase.

For the purpose of the present analysis, it is sufficient to retain only the terms in Eq.\,\eqref{eq:CFULL} that depend on $\ln \hat M_I$ through Eq.\,\eqref{eq:C1ij}, since these terms affect the two-zero minor structure. The SM radiative corrections in the EFT do not generate additional contributions that modify this structure. 
Therefore, we evaluate the neutrino mass using $C^{(\MSB)}(\mu)$. 
Furthermore, we reduce the boundary condition in Eq.\,\eqref{eq:C1ij} by omitting terms that are irrelevant to the two-zero minor structure. The reduced coefficient is then given by
\begin{align}
C^{(\MSB,\mathrm{red})}(\mu_M)
&=
C^{(0)}
+\frac{1}{2}\bigl(\theta^{L(\mathrm{red})}\bigr)^T C^{(0)}
+
\frac{1}{2}C^{(0)} \theta^{L\mathrm{(red)}}\cr
&\phantom{=}-\frac{1}{32\pi^2}\qty[
 C^{(0)} \hat{\kappa}^*
 \qty(\ln\frac{\mu_M^2}{\hat{M}^2})
 \hat{\kappa}^T
+ \hat{\kappa}
 \qty(\ln\frac{\mu_M^2}{\hat{M}^2})
 \hat{\kappa}^\dagger C^{(0)}
]\cr
&\phantom{=}- 
\frac{1}{64\pi^2}
\qty[3(g^2 + g^{\prime 2})+8\lambda_H]
\hat{\kappa} \hat{M}^{-1}
\qty(\ln\frac{\mu_M^2}{\hat{M}^2})
\hat{\kappa}^T \, ,
\label{eq:CMSBred}
\end{align}
where $\theta^{L\mathrm{(red)}}$ denotes the reduced form of Eq.\,\eqref{eq:thetaLII}, obtained by omitting the $+5/8$ term.

Notice that the above results can also be extracted from the general matching conditions for the type-I seesaw mechanism given in Ref.\,\cite{Zhang:2021jdf}, and are consistent with them. The contribution in the last line agrees with $\delta M_L$ given in Refs.\,\cite{Grimus:2002nk,AristizabalSierra:2011mn}, which provides the dominant one-loop correction to the neutrino masses. The first two lines arise from diagrams involving the charged-lepton self-energy and the charged-lepton Yukawa coupling with right-handed neutrinos. These contributions also affect the two-zero minor structure.

\section{Results}
\label{sec:analysis}
Taking into account the radiative corrections derived in the previous section, 
we perform a parameter fit to the neutrino observables. 
For the seesaw parameters, $M$ and $\kappa$, we numerically minimize 
the effective chi-square function $\mathit{\Delta} \chi^2_{\mathrm{eff}}$ defined in 
Appendix~\ref{app:chi2}
based on NuFIT~6.1 results~\cite{Esteban:2024eli}.
We also impose
\begin{align}
  \tr\qty[\kappa^\dagger \kappa] \le \pi
\end{align}
as a perturbativity condition.
For the matching scale, we choose the geometric mean of the right-handed 
neutrino masses,
\begin{align}
    \mu_M = ( \hat{M}_1 \hat{M}_2 \hat{M}_3 )^{1/3}\, .
\end{align}
The Wilson coefficient determined by the matching condition in
Eq.\,\eqref{eq:CMSBred} is then evolved down to the weak scale 
using the RG equation, and the neutrino mass matrix is
evaluated at that scale.

\begin{figure}[tbp]
    \centering
    \includegraphics[width=0.5\linewidth]{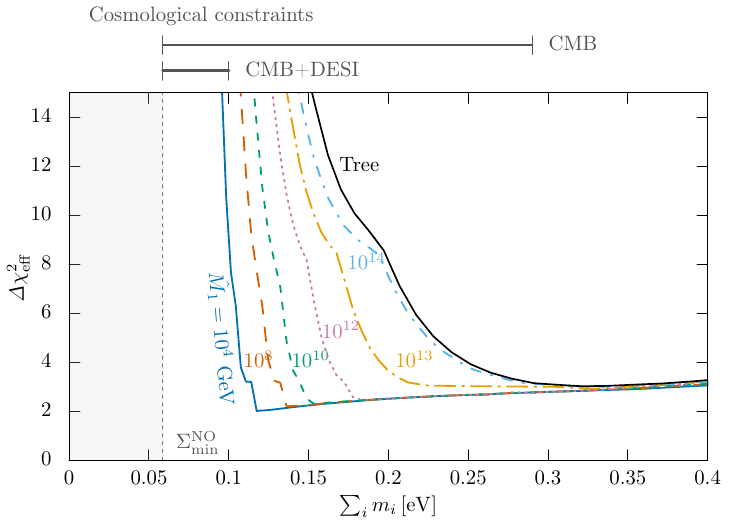}
    \caption{
    The effective chi-square, $\mathit{\Delta}\chi_\mathrm{eff}^2$, as a function of the total neutrino mass $\sum_i m_i$.
    The colored curves show the one-loop results for different lower bounds on $\hat{M}_1$, while the black solid curve shows the tree-level result.
    }
    \label{fig:chi2-mtot}
\end{figure}

In Fig.~\ref{fig:chi2-mtot}, we show $\mathit{\Delta}\chi_\mathrm{eff}^2$ as a function of the
total neutrino mass, $\sum_i m_i$.
The curve labeled ``Tree'' shows the result obtained with the exact
two-zero minor constraints
in Eq.\,\eqref{eq:minors}.  This corresponds to the analysis of
Ref.\,\cite{Ibe:2025rwk}, updated with the latest neutrino oscillation data.
The colored curves show the results including the one-loop corrections for
several lower bounds on the lightest right-handed-neutrino mass,
$\hat M_1$.

The figure shows that, for a given value of the total neutrino mass, the
chi-square value is significantly reduced for small values of $\hat M_1$
compared with the case of the exact two-zero minor constraints.
In fact, for the exact two-zero minor case,  $\mathit{\Delta}\chi_\mathrm{eff}^2$ becomes large for
$\sum_i m_i\lesssim 0.25\,\mathrm{eV}$.
For the case $\hat{M}_1 = 10^{4}$\,GeV, on the other hand, 
the total neutrino mass can be as low as
$\sum_i m_i = 0.12$\,eV.

This behavior can be traced back to the fact that the deviation from the
two-zero minor structure is induced by the $\ln \hat M$-dependent terms,
which are not proportional to the identity matrix in the right-handed-neutrino
mass basis.
Therefore, a larger hierarchy among the right-handed-neutrino masses leads
to a larger deviation from the tree-level two-zero minor structure, thereby
relaxing the constraints imposed by the exact two-zero minor conditions.

In Tab.\,\ref{tab:input-parameters}, we show the benchmark parameter sets
that minimize $\mathit{\Delta}\chi_\mathrm{eff}^2$ for several choices of the lower
bound on the lightest right-handed neutrino mass.
Here, we work in the flavor basis in which only right-handed neutrino mass parameter $M_{ee}$ 
in Eq.\,\eqref{eq:mu tau param}
carries a complex
phase, as discussed in Sec.\,\ref{sec:Radiative Correction on Two-Zero}.
The corresponding neutrino observables are summarized in
Tab.\,\ref{tab:observables}.

From these benchmark points, we see that large values of
$\epsilon_{\mu\mu}=\order{10^{-1}}$ are obtained where  $\kappa_\mu$ and
$\hat M_1$ are 
small.
As discussed around Fig.\,\ref{fig:r31-r21-epsilon}, such an
$\order{10^{-1}}$ deviation from the two-zero minor structure relaxes the
lower bound on the total neutrino mass.
Thus, the improvement in $\mathit{\Delta}\chi_\mathrm{eff}^2$ can be understood as a
consequence of the radiatively induced departure from the exact two-zero minor constraints.

In the exact two-zero minor case, the lower bound on the total neutrino
mass is in strong tension with cosmological constraints~\cite{Ibe:2016dir}.  
Once radiative
corrections are included, however, we find that this lower bound is
considerably reduced.  Comparing our result with the Bayesian 95\%
posterior upper limits
\begin{align}
\sum_i m_i <
\begin{cases}
0.29~{\rm eV}\ ,
& \text{Planck PR4 CMB+lensing, without BAO}\ , \\[2mm]
0.10~{\rm eV}\ ,
& \text{DESI DR2 BAO/DR1 full-shape + CMB}\ ,
\end{cases}
\end{align}
we find that the tension between the model prediction and cosmological
constraints is relaxed.
Here, the first limit is obtained by reweighting the public Planck PR4
\texttt{hlpTTTEEE\_lowT\_lolE\_lensing\_Mnu} chain of
Ref.\,\cite{Tristram:2023haj}, while the second corresponds to the DESI
95\% upper limit on the lightest neutrino mass,
$m_{\rm lightest}<0.023\,{\rm eV}$, mapped to $\sum_i m_i$ for the
normal-ordering spectrum~\cite{Elbers:2025vlz}.  In our analysis, we 
restrict the cosmological posterior to the normal-ordering region,
\begin{align}
\sum_i m_i \geq \Sigma_{\rm min}^{\rm NO}
\simeq 0.059~{\rm eV}\ ,
\end{align}
while keeping the original cosmological prior on $\sum_i m_i$.

As a result, the $\mathrm{U}(1)_{L_\mu-L_\tau}$ model still exhibits some
tension with neutrino-mass bounds obtained from cosmological data sets that
include BAO measurements.  On the other hand, if we adopt the more
conservative CMB-only constraint, the model remains broadly consistent
with current cosmological observations.  It should also be noted that the
neutrino-mass bounds inferred from BAO data are subject to ongoing
discussions; see, for example, Ref.\,\cite{Naredo-Tuero:2024sgf}.

\begin{table}[t]
\centering
\scriptsize
\setlength{\tabcolsep}{3pt}
\renewcommand{\arraystretch}{1.15}
\caption{Input parameters for the benchmark points.
At the benchmark point $\mathrm{BP}_i$, the lightest right-handed neutrino mass is
chosen as $\hat M_1 = 10^i\,{\rm GeV}$.
For the matching scale, we choose the geometric mean of the right-handed 
neutrino masses,
$ \mu_M = ( \hat{M}_1 \hat{M}_2 \hat{M}_3 )^{1/3}$.
}
\label{tab:input-parameters}
\begin{tabular}{c ccccc ccc}
\hline
BP & \multicolumn{5}{c}{Mass parameters in GeV unit} & \multicolumn{3}{c}{Yukawa parameters}\\
\cline{2-6}\cline{7-9}
 & $\mathrm{Re}\,M_{ee}$ & $\mathrm{Im}\,M_{ee}$ & $M_{e\mu}$ & $M_{e\tau}$ & $M_{\mu\tau}$ & $\kappa_{e}$ & $\kappa_{\mu}$ & $\kappa_{\tau}$ \\
\hline
$\mathrm{BP}_{4}$ & $-8.6027\times 10^{13}$ & $-6.4354\times 10^{12}$ & $2.4387\times 10^{8}$ & $4.6099\times 10^{13}$ & $3.8299\times 10^{8}$ & $0.38803$ & $2.1812\times 10^{-6}$ & $0.38093$ \\
$\mathrm{BP}_{8}$ & $-3.0506\times 10^{14}$ & $-1.6715\times 10^{13}$ & $2.9121\times 10^{10}$ & $1.7680\times 10^{14}$ & $8.5708\times 10^{10}$ & $0.78084$ & $1.9580\times 10^{-4}$ & $1.0513$ \\
$\mathrm{BP}_{10}$ & $-5.7683\times 10^{13}$ & $-2.6112\times 10^{12}$ & $8.6700\times 10^{10}$ & $7.9599\times 10^{13}$ & $9.3500\times 10^{11}$ & $0.36637$ & $0.001736$ & $1.411$ \\
$\mathrm{BP}_{12}$ & $-7.2592\times 10^{13}$ & $-2.9298\times 10^{12}$ & $4.5417\times 10^{11}$ & $6.6771\times 10^{13}$ & $7.5398\times 10^{12}$ & $0.44891$ & $0.013526$ & $1.7146$ \\
$\mathrm{BP}_{14}$ & $-2.5704\times 10^{14}$ & $2.1868\times 10^{12}$ & $3.5901\times 10^{13}$ & $8.4166\times 10^{13}$ & $1.0607\times 10^{14}$ & $1.1931$ & $0.48765$ & $1.2167$ \\
\hline
\end{tabular}
\end{table}

\begin{table}[t]
\centering
\scriptsize
\setlength{\tabcolsep}{5pt}
\renewcommand{\arraystretch}{1.15}
\caption{Neutrino observables for the benchmark points.}
\label{tab:observables}
\begin{tabular}{c c c c c c c}
\hline
Observable & NuFIT~6.1 & $\mathrm{BP}_{4}$ & $\mathrm{BP}_{8}$ & $\mathrm{BP}_{10}$ & $\mathrm{BP}_{12}$ & $\mathrm{BP}_{14}$ \\
\hline
$\sum_i m_i\,[\mathrm{eV}]$ & - & $0.11748$ & $0.13619$ & $0.15328$ & $0.18302$ & $0.32093$ \\
$\mathit{\Delta} m_{21}^2\,[10^{-5}\,\mathrm{eV}^2]$ & $7.537^{+0.094}_{-0.10}$ & $7.5283$ & $7.532$ & $7.5351$ & $7.5476$ & $7.5386$ \\
$\mathit{\Delta} m_{31}^2\,[10^{-3}\,\mathrm{eV}^2]$ & $2.511^{+0.021}_{-0.020}$ & $2.5099$ & $2.505$ & $2.505$ & $2.5053$ & $2.505$ \\
$\sin^2\theta_{12}$ & $0.3088^{+0.0067}_{-0.0066}$ & $0.30851$ & $0.30801$ & $0.30773$ & $0.30622$ & $0.30868$ \\
$\sin^2\theta_{23}$ & $0.470^{+0.017}_{-0.014}$ & $0.46498$ & $0.465$ & $0.465$ & $0.465$ & $0.46482$ \\
$\sin^2\theta_{13}$ & $0.02248^{+0.00055}_{-0.00049}$ & $0.022293$ & $0.022289$ & $0.022297$ & $0.022293$ & $0.022226$ \\
$\delta_{\rm CP}\,[{}^\circ]$ & $212^{+26}_{-36}$ & $254.6$ & $258.2$ & $260.37$ & $263.7$ & $281.02$ \\
$m_{\beta}\,[\mathrm{meV}]$&-&$30$&$38$&$44$&$55$&$103$\\
$m_{\beta\beta}\,[\mathrm{meV}]$&-&$24$&$33$&$40$&$53$&$94$\\
\hline
$\mathrm{Re} \, \epsilon_{\mu\mu}$ &-&$0.15$&$0.18$&$0.19$&$0.19$&$1.1\times 10^{-3}$\\
$\mathrm{Im} \, \epsilon_{\mu\mu}$ &-&$-0.22$&$-0.17$&$-0.14$&$-8.2\times 10^{-2}$&$8.8\times 10^{-4}$\\
$\mathrm{Re} \, \epsilon_{\tau\tau}$ &-&$9.3\times 10^{-5}$&$3.4\times 10^{-3}$&$1.2\times 10^{-3}$&$2.0\times 10^{-3}$&$9.6\times 10^{-4}$\\
$\mathrm{Im} \, \epsilon_{\tau\tau}$ &-&$6.2\times 10^{-5}$&$-3.3\times 10^{-3}$&$-9.3\times 10^{-4}$&$-9.5\times 10^{-4}$&$7.6\times 10^{-4}$\\
\hline
\end{tabular}
\end{table}

In the exact two-zero minor case, where the lower bound on the total neutrino mass is relatively high, the constraint from neutrinoless 
double-beta decay also plays an important role.
The decay
rate is proportional to the square of the effective Majorana mass,
\begin{align}
\label{eq:mbb_def}
m_{\beta\beta}
=
\left|
\sum_i U_{ei}^2 m_i
\right|\ ,
\end{align}
which is constrained by KamLAND-Zen~\cite{KamLAND-Zen:2024eml} as
\begin{align}
m_{\beta\beta}
<
(28\text{--}122)\,\mathrm{meV}
\qquad
(90\%~\mathrm{C.L.})\ ,
\end{align}
depending on the nuclear matrix element.

Once radiative corrections are included, however, the predicted value of
$m_{\beta\beta}$ can be reduced, as shown in
Tab.\,\ref{tab:observables}.
The values obtained in our benchmark points
are below the current KamLAND-Zen bound.  Therefore, the constraint from
neutrinoless double-beta decay is significantly relaxed by radiative
corrections.

We also quote 
the constraint
on the 
effective electron neutrino mass in $\beta$ decay,
\begin{align}
\label{eq:mbeta}
m_\beta^2
=
\sum_i |U_{ei}|^2 m_i^2\ .
\end{align}
The KATRIN experiment currently sets the upper bound
\begin{align}
m_\beta < 450\,{\rm meV}
\qquad
(90\%~{\rm C.L.})\, .
\end{align}
The predicted values 
in 
Tab.\,\ref{tab:observables}
are smaller than the upper limit for the benchmark points.

We should also note that the present scan demonstrates the existence of
parameter points with small $\mathit{\Delta}\chi^2_{\mathrm{eff}}$, but does not quantify
the volume of such regions in the high-scale parameter space.  The low-$\chi^2$
points may involve nontrivial correlations among the right-handed-neutrino
masses, Yukawa couplings, and CP phases, because the loop correction must
break the two-zero-minor structure in a suitable direction.  While this does
not affect the frequentist interpretation adopted here, it may be relevant
from a Bayesian point of view, where the prior volume and the evidence are
sensitive to such correlations.

Finally, let us comment on the theoretical uncertainty associated with the
radiative corrections.  In the present analysis, all right-handed neutrinos are
decoupled at a common matching scale $\mu_M$, and the large logarithms
associated with hierarchical right-handed-neutrino masses are not resummed.
As a result, the prediction at a fixed point in the high-scale parameter space
has a residual large dependence on $\mu_M$, especially for hierarchical spectra.

This point-by-point scale dependence should not, however, be directly
interpreted as an uncertainty in the existence of the viable parameter region.
The main purpose of the present analysis is to estimate the qualitative impact
of radiative corrections on the two-zero minor structure.  As shown in
Appendix~\ref{app:matching_scale}, 
we have repeated the parameter scan for
several choices of $\mu_M$ and found that the low-$\mathit{\Delta}\chi^2_{\rm eff}$
region at smaller $\sum_i m_i$ persists.  Thus, although a precision analysis
of individual benchmark points would require sequential decoupling of the
right-handed neutrinos together with the running and matching across their
mass thresholds~\cite{Antusch:2002rr,Antusch:2005gp}, the qualitative
conclusion of this work is not sensitive to the choice of the common matching
scale.

\section{Conclusions}
\label{sec:conclusions}

In this paper, we revisited the constraints from the two-zero-minor structure by taking the 
minimal $\Umt$ model as an example with
one-loop radiative corrections to the seesaw-induced neutrino mass matrix.
At tree level, the $\Umt$ symmetry enforces the two-zero minor structure of the light-neutrino mass matrix, which leads to strong correlations among the neutrino oscillation parameters, CP phases, and the absolute neutrino mass scale.
In particular, for the normal ordering, the tree-level two-zero-minor conditions favor a quasi-degenerate neutrino mass spectrum with rather large total neutrino mass, 
and therefore lead to strong tension with the cosmological constraints on the total neutrino mass.
We computed the one-loop corrections induced by the interactions relevant to the seesaw mechanism
and  performed a statistical analysis based on the 
latest neutrino oscillation data. 
We found that the one-loop corrections significantly relax the rigid tree-level constraint on the total neutrino mass.
As a result,
we find $\mathit{\Delta}\chi_\mathrm{eff}^2$ is reduced for a given value of 
the total neutrino mass
compared with the tree-level analysis.
This shows that the two-zero minor prediction of the minimal $\Umt$ model is not completely robust against radiative corrections.

In this paper, we have considered, as threshold corrections, only the
radiative corrections induced by the interactions between the Standard
Model fields and the right-handed neutrinos.  As pointed out in
Ref.\,\cite{Ibe:2016dir} (see also Ref.\,\cite{Ibarra:2026max}), however, interactions with the $\mathrm{U}(1)_{L_\mu-L_\tau}$
gauge boson and the $\mathrm{U}(1)_{L_\mu-L_\tau}$ Higgs sector can also
break the two-zero minor structure.  We leave a detailed study of these
effects for future work.
It would also be interesting to apply the analysis developed in this paper
to other $\mathrm{U}(1)_{L_\alpha-L_\beta}$ models ($\alpha,\beta=e,\mu,\tau$), as well as to other models that predict special neutrino mass textures or minor structures at
tree level.

\section{Acknowledgments}
This work is supported by Grant-in-Aid for Scientific Research from the Ministry of Education, Culture, Sports, Science, and Technology (MEXT), Japan,  24K23938, 25H00644, 26K07102
(M.I.), 26K00719 and 26K22326 (S.S.) and by World Premier International Research Center Initiative (WPI),
MEXT, Japan.

\appendix
\section{Statistical Treatment of Neutrino Oscillation Data}
\label{app:chi2}

Neutrino oscillations are described by six parameters,
\begin{gather}
\label{eq:oscillation parameters}
\mathit{\Delta} m_{21}^2\ , \quad
\mathit{\Delta} m_{3\ell}^2\ , \cr
\sin^2\theta_{12}\ , \quad
\sin^2\theta_{13}\ , \quad
\sin^2\theta_{23}\ , \quad
\delta_{\mathrm{CP}}\ .
\end{gather}
Although NuFIT~6.1 provides comprehensive $\mathit{\Delta}\chi^2$ data for these
parameters, the full six-dimensional $\mathit{\Delta}\chi^2$ function is not publicly
available.  Instead, NuFIT provides $\mathit{\Delta}\chi^2$ tables marginalized over
subsets of one, two, or three variables, which we denote by
$\mathit{\Delta}\chi^2_{\mathrm{1D}}$, $\mathit{\Delta}\chi^2_{\mathrm{2D}}$, and
$\mathit{\Delta}\chi^2_{\mathrm{3D}}$, respectively.  In these tables, all parameters
other than the displayed ones are minimized over.  For the normal neutrino mass
ordering, the minimum value of the delta chi-square is zero.

For the analysis of the allowed region in the $(r_{31},r_{21})$ plane, we
introduce an effective chi-square function.  Since the full
four-dimensional chi-square data in the parameter space
$(s_{12}^2,s_{13}^2,s_{23}^2,\delta_{\mathrm{CP}})$ are not publicly
available, we define
\begin{align}
\label{eq:chisquare mixing}
\mathit{\Delta}\chi^2_{\mathrm{mix}}
(s_{12}^2,s_{13}^2,s_{23}^2,\delta_{\mathrm{CP}})
=
\max\left[
\mathit{\Delta}\chi^2_{\mathrm{2D}}(x,y),
\,
\mathit{\Delta}\chi^2_{\mathrm{1D}}(z)
\right]\ ,
\end{align}
where $x$, $y$, and $z$ collectively denote the relevant oscillation
parameters.  The maximum is taken over all available one- and
two-dimensional NuFIT tables involving the mixing parameters and the Dirac
CP phase.
We should note that $\mathit{\Delta}\chi^2_{\mathrm{mix}}$ may underestimate the full
chi-square in the four-dimensional parameter space
$(s_{12}^2,s_{13}^2,s_{23}^2,\delta_{\mathrm{CP}})$.  Therefore, the
constraints on $(r_{31},r_{21})$ obtained from this effective chi-square
should be regarded as conservative.

For the constraint on the total neutrino mass, on the other hand, we use
the effective chi-square function
\begin{align}
\label{eq:chisquare effective}
\mathit{\Delta}\chi^2_{\mathrm{eff}}
\left(
\mathit{\Delta} m_{21}^2,
\mathit{\Delta} m_{3\ell}^2,
s_{12}^2,
s_{13}^2,
s_{23}^2,
\delta_{\mathrm{CP}}
\right)
=
\max\left[
\mathit{\Delta}\chi^2_{\mathrm{3D}}
\left(
\mathit{\Delta} m_{3\ell}^2,
s_{23}^2,
\delta_{\mathrm{CP}}
\right),
\,
\mathit{\Delta}\chi^2_{\mathrm{2D}}(x,y),
\,
\mathit{\Delta}\chi^2_{\mathrm{1D}}(z)
\right].
\end{align}
The maximum is again taken over all available one- and two-dimensional
NuFIT tables, together with the three-dimensional table for
$(\mathit{\Delta}m_{3\ell}^2,\sin^2\theta_{23},\delta_{\mathrm{CP}})$.
This effective chi-square may also underestimate the full chi-square in
the six-dimensional parameter space
$(
\mathit{\Delta} m_{21}^2,
\mathit{\Delta} m_{3\ell}^2,
s_{12}^2,
s_{13}^2,
s_{23}^2,
\delta_{\mathrm{CP}}
)$.
Therefore, the lower bound on the total neutrino mass obtained in our
analysis should be regarded as conservative.

For the tree-level two-zero-minor case, the numerical analysis in Sec.\,\ref{sec:analysis} gives a minimum value of $\mathit{\Delta}\chi^2_{\rm eff}\simeq 3$.  Since
$\mathit{\Delta}\chi^2_{\mathrm{eff}}$ is constructed from marginalized NuFIT tables, it is
expected to underestimate the full chi-square in the six-dimensional
oscillation-parameter space.  Therefore, the corresponding likelihood-ratio
test statistic in the full parameter space is expected to be at least of this
size.  If Wilks' theorem with one effective degree of freedom is applied, this
corresponds to a $p$-value of about $8\%$, or equivalently to a disfavoring of
the tree-level two-zero-minor hypothesis at about the $92\%$ confidence level
relative to the general normal-ordering case.  This interpretation should be
regarded as indicative, because the publicly available NuFIT data do not allow
us to construct the full likelihood.

For the one-loop analysis, on the other hand, the two-zero minor structure
is no longer exact once the one-loop corrections in
Eq.\,\eqref{eq:CMSBred} are included.  The breaking of the two-zero minor
structure introduces additional parameter dependences in the predicted
oscillation observables.  Moreover, the model parameter space is subject
to theoretical constraints, such as lower bounds on the right-handed
neutrino masses and upper bounds on the sizes of the $\kappa$'s.  For
these reasons, Wilks' theorem cannot be applied in a straightforward
manner, and the statistical interpretation of
$\mathit{\Delta}\chi^2_{\mathrm{eff}}$ is not simple.

Nevertheless, a small value of $\mathit{\Delta}\chi^2_{\mathrm{eff}}$ is still
expected to indicate a good fit between the model prediction and the
oscillation data.  What is difficult with the currently available NuFIT
data is to assign a precise frequentist interpretation to
$\mathit{\Delta}\chi^2_{\mathrm{eff}}$, such as a quantitative likelihood-ratio
test.  We therefore use $\mathit{\Delta}\chi^2_{\mathrm{eff}}$ as a practical
measure of the compatibility between the model prediction and the
oscillation data, rather than as a statistic with a definite number of
degrees of freedom.

\section{Canonical and Diagonal Basis of Charged Leptons}
\label{app:diagonal lepton}
The UV contributions to the renormalized two-point function of the charged leptons and to the charged-lepton Yukawa vertex function appear in 
the 1PI effective Lagrangian, 
\begin{align}
    \mathcal{L}_{\mathrm{1PI}}= L^{\dagger}_i
    i\partial_\mu\bar{\sigma}^\mu 
    \qty(\delta_{ij}-\Sigma^L_{ij})L_j
    - H^\dagger \bar{E}_i y_i
     \qty(\delta_{ij} +\Gamma_{H\bar{E}_iL_j})L_j\
    \ .
\end{align}
Then, the canonically normalized 
charged lepton $L'$ and Higgs doublet $H'$ are obtained as,
\begin{align}
    L_i =  \qty(\delta_{ij}+\frac{1}{2}\Sigma^L_{ij})L_j' \ ,
    \quad H =   \qty(1+\frac{1}{2}\Sigma^H)H' \ ,
\end{align}
which induces the effective Yukawa coupling 
\begin{align}
    \mathcal{L}_\mathrm{Yukawa}
    &= - H'^\dagger\bar{E}_iy_i\qty(\delta_{ij}+\Delta_{ij})L'_j +\mathrm{h.c.} \\
    \Delta_{ij}&:= \frac{1}{2}\Sigma^L_{ij} + \frac{1}{2} \Sigma^H \delta_{ij} + \Gamma_{H\bar{E}_iL_j}\ .
\end{align}

Finally, by using unitary rotations, 
\begin{align}
    L_i' &= (U_L)_{ij} \hat{L}_j \simeq \qty(\delta_{ij}
    +\frac{1}{2}\theta^L_{ij})\hat{L}_j\ ,\\
      \bar{E}_i' &= (U_E)_{ij} \hat{\bar{E}}_j \simeq \qty(\delta_{ij}
    +\frac{1}{2}\theta^E_{ij})\hat{\bar{E}}_j\ ,
\end{align}
we obtain the flavor basis where the 
charged-lepton Yukawa couplings are diagonal after radiative correction.
Here, the anti-hermitian matrices $\theta^{L,E}$ are given by
\begin{align}
\frac{1}{2}    \theta^L_{ij} = \begin{cases}
        -\displaystyle{\frac{y_i^2+y_j^2}{y_{i}^2-y_{j}^2}}
    \Delta_{ij}& i\neq j\\
    0 & i = j 
    \end{cases}\ , \qquad
  \frac{1}{2}        \theta^E_{ij} = \begin{cases}-
\displaystyle{    \frac{2y_iy_j}{y_{i}^2-y_{j}^2}
    \Delta_{ij}^*
    } & i\neq j\\
    0 & i = j 
    \end{cases}\ .
\end{align}
By noting the hierarchy of the charged-lepton Yukawa coupling, we find 
\begin{align}
\label{eq:thetaL}
    \frac{1}{2}\theta_L\simeq 
    \mqty(0&\Delta_{e\mu}&\Delta_{e\tau}\\
    -\Delta^*_{e\mu}&0&\Delta_{\mu\tau}\\
    -\Delta^*_{e\tau}&-\Delta^{*}_{\mu\tau}&0
    ) \ ,\qquad \theta_E \simeq 0 \ . 
\end{align}
As a result, we find the canonically normalized charged-lepton doublet is given by
\begin{align}
\label{eq:canonical diagonal L}
    L_i \simeq \qty(\delta_{ij}+\frac{1}{2}
    \qty(\Sigma_{ij}^L + \theta^L_{ij}))\hat{L}_j\ .
\end{align}
For use in the main text, we define $\hat{\Sigma}^L$ as follows,
\begin{align}
\hat{\Sigma}^L_{ij}:= 
     \qty(\Sigma^L_{ij} + \theta^L_{ij}) \ .
    \label{eq:charged-lepton rotation}
\end{align}

By substituting the results in Sec.\,\ref{sec:Radiative Correction on Two-Zero}, 
\begin{align}
    \Sigma^H=  -\frac{1}{16\pi^2}
   \tr[\kappa^* \qty(\ln\frac{\mu^2}{\hat{M}^2} + \frac{1}{2})\kappa^T]\ ,
\end{align}
\begin{align}
 \Sigma^L = 
    -\frac{1}{32\pi^2} \hat{\kappa}^*\qty(\ln\frac{\mu^2}{\hat{M}^2}+\frac{3}{2})\hat{\kappa}^
    T\ ,
\end{align}
\begin{align}
\Gamma_{H\bar{E}L} =
\frac{1}{16\pi^2} 
\hat{\kappa}^*\qty(\ln \frac{\mu^2}{\hat{M}^2}+1)\hat{\kappa}^T \ ,
\end{align}
we obtain,
\begin{align}
\label{eq:Delta}
    \Delta
    = \frac{1}{16\pi^2} \hat{\kappa}\qty( \frac{3}{4}\ln \frac{\mu_M^2}{\hat{M}^2} + \frac{5}{8})\hat{\kappa}^\dagger 
    - \frac{1}{32\pi^2}\mathrm{Tr} \Bigg[\hat{\kappa}\Bigg(\ln \frac{\mu_M^2}{\hat{M}^2} + \frac{1}{2}\Bigg)\hat{\kappa}^\dagger\Bigg] \ .
\end{align}
Notice that the second term which comes from $\Sigma^H$ does not affect $\theta^L$ as it is flavor diagonal.

\section{Dependence on the Matching Scale}
\label{app:matching_scale}

\begin{figure}[t]
  \centering
  \includegraphics[width=0.5\linewidth]{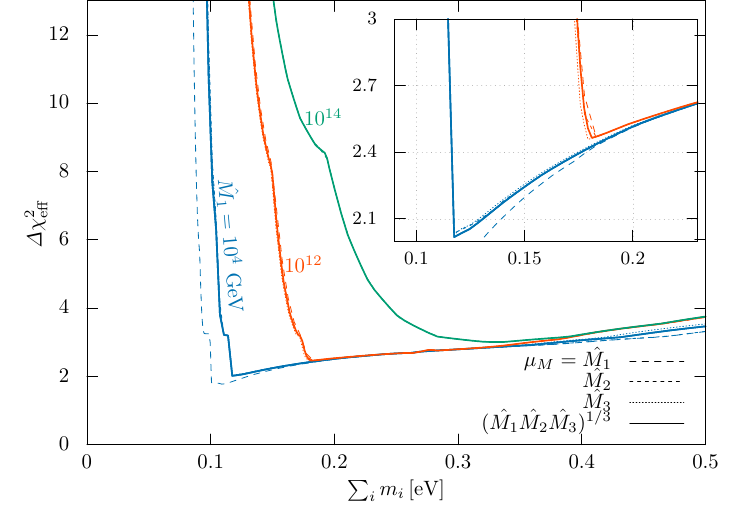}
  \caption{
    Dependence of $\Delta\chi^2_{\rm eff}$ on the common matching scale
    $\mu_M$.  The analysis is repeated for
    $\mu_M=\hat M_1,\hat M_2,\hat M_3$ and
    $\mu_M=(\hat M_1\hat M_2\hat M_3)^{1/3}$.
    The colors denote the lower bound imposed on the lightest right-handed
    neutrino mass, as in Fig.\,\ref{fig:chi2-mtot}.  The inset magnifies the
    low-$\Delta\chi^2_{\rm eff}$ region.  Although the curves show a residual
    matching-scale dependence, the one-loop improvement at low
    $\sum_i m_i$ is stable under these variations.
  }
  \label{fig:matching_scale_dependence}
\end{figure}

In the main text, the one-loop-corrected Wilson coefficient is evaluated by
integrating out the three right-handed neutrinos at a common matching scale
$\mu_M$.  Since the reduced matching condition in Eq.\,\eqref{eq:CMSBred} contains
logarithms of the form ${\ln}(\mu_M^2/\hat M_I^2)$, a residual dependence on
$\mu_M$ remains in a fixed-order calculation when the right-handed neutrino
spectrum is hierarchical.  This dependence should be regarded as an estimate
of higher-order effects associated with the common-scale matching prescription.
A complete treatment for a strongly hierarchical spectrum would require
sequential decoupling of the right-handed neutrinos together with the
corresponding renormalization-group evolution between thresholds.

In the numerical analysis in Sec.\,\ref{sec:analysis}, we use
\begin{align}
  \mu_M = (\hat M_1 \hat M_2 \hat M_3)^{1/3}
\end{align}
as the default matching scale.  To estimate the size of the residual matching
scale dependence, we repeat the analysis for
\begin{align}
  \mu_M = \hat M_1\ ,\quad
  \hat M_2\ ,\quad
  \hat M_3\ ,\quad
  (\hat M_1 \hat M_2 \hat M_3)^{1/3}\ .
\end{align}
The result is shown in Fig.\,\ref{fig:matching_scale_dependence}.  The same
definition of $\mathit{\Delta}\chi^2_{\rm eff}$ as in Appendix~\ref{app:chi2} is used.

We find that the quantitative value of $\mathit{\Delta}\chi^2_{\rm eff}$ is mildly
affected by the choice of $\mu_M$, as expected from the presence of large
logarithms for hierarchical right-handed neutrino masses.  Nevertheless, the
qualitative behavior is stable.  In particular, for the most hierarchical case
with $\hat M_1=10^4\,\mathrm{GeV}$, the one-loop correction still allows the
minimum of $\mathit{\Delta}\chi^2_{\rm eff}$ to occur at a substantially smaller value
of the total neutrino mass than in the tree-level two-zero-minor case.  The
improvement of the fit at low $\sum_i m_i$ therefore does not rely on a special
choice of the common matching scale.

This scale-dependence check supports the conclusion that the radiatively
induced breaking of the two-zero-minor structure can relax the lower bound on
the total neutrino mass.  The residual $\mu_M$ dependence is a limitation of the
fixed-order common-scale matching used in this work, but it does not change the
main conclusion.

\bibliographystyle{apsrev4-1}
\bibliography{ref}

\end{document}